\documentstyle[pra,aps]{revtex}
\begin{document}
\title{Common Molecular Dynamics Algorithms Revisited: Accuracy and Optimal Time
Steps of St\"ormer-Leapfrog Integrators}
\author{Alexey K. Mazur}
\address{Laboratoire de Biochimie Th\'eorique, CNRS UPR9080 Institue de
Biologie Physico-Chimique 13, rue Pierre et Marie Curie, Paris,75005, France}

\date{\today}
\maketitle

\begin{abstract}
The St\"ormer-Verlet-leapfrog group of integrators commonly used in molecular
dynamics simulations has long become a textbook subject and seems to have
been studied exhaustively. There are, however, a few striking effects in
performance of algorithms which are well-known but have not received adequate
attention in the literature. A closer view of these unclear observations
results in unexpected conclusions. It is shown here that contrary to the
conventional point of view, the leapfrog scheme is distinguished in this
group both in terms of the order of truncation errors and the conservation of
the total energy. In this case the characteristic square growth of
fluctuations of the total energy with the step size, commonly measured in
numerical tests, results from additional interpolation errors with no
relation to the accuracy of the computed trajectory. An alternative procedure
is described for checking energy conservation of leapfrog-like algorithms
which is free from interpolation errors. Preliminary tests on a
representative model system suggest that standard step size values used at
present are lower than necessary for accurate sampling.\end{abstract}

\section{Introduction}
The well-known group of integrators for the Newton's equations comprising
Verlet \cite{Verlet:67}, leapfrog \cite{Hockney:81}, velocity Verlet
\cite{Swope:82}, and Beeman \cite{Beeman:76} methods play a central role in
the classical methodology of molecular dynamics. Due to their simplicity and
exceptional stability they proved to be the best choice for long time - large
step size calculations and they are now employed in the great majority of
simulations of large systems. All or some of these methods are always
described in detail and compared in any modern textbook
\cite{Hockney:81,Allen:87,Haile:92}. It is well-known that they are based
upon a ninety-year-old St\"ormer time-centered difference approximation of
accelerations \cite{Venneri:87,Gear:71}, that given appropriate initial
conditions they produce the same trajectory in coordinate space and that,
therefore, they all represent variations of a single algorithm.

Despite this long prehistory, the performance of these algorithms still
attracts remarkable attention, first, because of their practical importance
and, second, because understanding of the origin of their exceptional
properties may help to develop even better algorithms. In particular, their
stability has been attributed to the correspondence between the order of the
finite difference equation and its analytical analog \cite{Hockney:81}, to
their time reversibility \cite{Hockney:81,Tuckerman:92,Toxvaerd:93}, to
oscillating fundamental solutions \cite{Janzen:84} and to their symplectic
property \cite{Tuckerman:92,Auerbach:91,Toxvaerd:93a}. A hallmark of these
method is a square power growth of global errors with the step size, which is
invariably observed with different testing techniques and can even be used
for debugging computer codes \cite{Allen:87}. Because of the low apparent
order of approximation they usually appear less accurate than other methods
in comparative studies, but, with large step sizes, when other algorithms
loose stability, the leapfrog scheme and its analogs still produce
trajectories with very low total energy drift and correct average static and
dynamic properties \cite{vanGunsteren:77,Macgowan:88,Fincham:86}.  These
observations are known and exploited for such a long time that it seems to
have been forgotten that actually they are quite unusual and that this
behavior still has no clear explanation.

The observed persistence of averages and a low drift of the total energy
might mean that the numerical trajectory computed with a large step size,
though inaccurate in a strict sense, holds well to the correct constant
energy hypersurface in phase space. It is well-known, however, that compared
with other integrators the St\"ormer-equivalent algorithms conserve the
instantaneous total energy rather poorly - its value already strongly
fluctuates at time steps well below the theoretical limit of stability.
Thus, it appears that the trajectory constantly deviates from the initial
hypersurface, but always finds a way back, which is rather striking because,
for example, the velocity Verlet algorithm is self-starting and,
consequently, keeps no memory of the preceding part of the trajectory. One
explanation to these observations follows from the general property of symplectic
integrators which, with a sufficiently small time step, generate
discretizations of exact trajectories corresponding to perturbed Hamiltonians
\cite{Auerbach:91}. This property, however, holds for small time steps only
and it does not explain why the leapfrog-equivalent algorithms are
distinguished among symplectic integrators as well.

The above contradiction may be settled if one assumes that the total energy
is actually conserved better than it appears. For a St\"ormer or a leapfrog
trajectory this is possible, in principle, because in these cases the total
energy is not a well-defined quantity. The original St\"ormer formula
\cite{Verlet:67,Allen:87} needs only coordinates and accelerations for
computing a trajectory and so, strictly speaking, velocities are not defined.
There are many methods to evaluate velocities and kinetic energies but they
all employ some finite difference approximations and, usually,
interpolations, which is often done implicitly, as in the case of the
velocity Verlet and Beeman algorithms \cite{Swope:82,Beeman:76}. It is
understood that these computations may introduce additional errors into the
computed total energy, but these errors are difficult to evaluate or to get
rid of and one cannot generally tell how large is their relative
contribution.

The initial goal of the present study was to check more accurately how well a
leapfrog trajectory holds to a constant-energy hypersurface, when only
approximations intrinsic to the algorithm play a role and there is no
influence of additional approximations and interpolations. As happens
sometimes, the solution of a practical task has lead to a comprehensive
``{\em ab initio}'' analysis of several related problems which have been
unrecognized or simply ignored in the literature. It is shown here that
contrary to the conventional point of view, the leapfrog scheme is more
accurate than other algorithms of this group, both in terms of the order of
the truncation errors and the conservation of the total energy. The observed
square growth of fluctuations of the total energy with the step size appears
to result from interpolations only and has no relation to the accuracy of the
computed trajectory. An alternative procedure is proposed for checking energy
conservation of leapfrog-like algorithms which is free from interpolation
errors. Testing on an example of protein dynamics confirms that the leapfrog
trajectory really manages to sample from a correct hypersurface in phase
space with larger step sizes than commonly recommended.

When a classical text-book subject is discussed it is difficult to maintain
the logic of argumentation without reproducing some well-known general
results. The author hopes, therefore, to be excused by experts if some of
the derivations in the text appear trivial.

\section{Results and Discussion}
We will consider Newton's equation for a particle of a unit mass
\begin{equation}
\ddot x=f\left(x\right)
\end{equation}
where the dot notation denotes time derivatives. The St\"ormer fourth order
algorithm (the order of the algorithm here and below refers to the order of
the truncation error) first introduced in this field by Verlet \cite{Verlet:67}
is
\begin{equation}
x_{n+1}=2x_n-x_{n-1}+a_nh^2
\end{equation}
where $x_n$ and $a_n$ are the coordinate and the acceleration at the nth time step
and $h$ is the step size. The leapfrog scheme popularized by Hockney and
Eastwood \cite{Hockney:81} is
{\mathletters\begin{eqnarray}
v_{n+\frac 12}=v_{n-\frac 12}+a_nh\\
x_{n+1}=x_n+v_{n+\frac 12}h
\end{eqnarray}}
where $v_{n+\frac 12}$ is the half-step velocity. The Swope et al.
\cite{Swope:82} formulation of the algorithm referred to as velocity Verlet is
{\mathletters\begin{eqnarray}
x_{n+1}=x_n+\left(v_n+a_n\frac h2\right)h\\
v_{n+1}=v_n+\left(a_n+a_{n+1}\right)\frac h2
\end{eqnarray}}
Finally, the Beeman method \cite{Beeman:76}
{\mathletters\begin{eqnarray}
x_{n+1}=x_n+v_nh+\left(\frac 23a_n-\frac 16a_{n-1}\right)h^2\\
v_{n+1}=v_n+\left(\frac 13a_{n+1}+\frac 56a_n-\frac 16a_{n-1}\right)h
\end{eqnarray}}
Instructive discussions of the relationships between these methods can be
found elsewhere \cite{Hockney:81,Allen:87,Toxvaerd:93,Tuckerman:93,Rodger:89}.
The algorithms defined by Eqs. (2)-(5)
significantly differ in the implicitly adopted point of view upon the
definition of the on-step velocities and, consequently, the total energy.
For the velocity Verlet and Beeman integrators the on-step velocity $v_n$,
kinetic energy $K_n$ and the total energy $E_n$ are well defined quantities. In
contrast, for both the leapfrog and the original St\"ormer algorithms no
complete trajectory in phase space is computed and one can choose between
different interpolation formulae for the calculation of on-step velocities
or kinetic energies. Because of the reasons given below it is most
convenient for us to take the second choice. Let us consider that our
trajectory is computed by the leapfrog integrator (3a,b). The coordinates
thus obtained automatically satisfy Eq. (2). The corresponding solution of
Eqs. (4a,b) is obtained if we employ the interpolation formula
\begin{equation}
v_n=\frac 12\left(v_{n-\frac 12}+v_{n+\frac 12}\right)
\end{equation}
and the solution of Eqs. (5a,b) with
\begin{equation}
v_n=\frac 16\left(2v_{n+\frac 12}+5v_{n-\frac 12}-v_{n-\frac 32}\right)
\end{equation}
It should be noted that since the Beeman algorithm employs coordinates and
velocities from several time steps it is equivalent to the leapfrog
algorithm with interpolation (7) only if its initial conditions match Eqs.
(3a,b).

When analytical trajectories are described it is common to use the term
``state'' to refer to a point in phase space. From any state the analytical
trajectory can be continued in a unique way in both time directions. For the
following discussion, it will be convenient to use a similar terminology
referring to numerical trajectories. Thus, a leapfrog state is defined as a
pair $\left(x_n,v_{n-1/2}\right)$, a velocity Verlet state as $\left(
x_n,v_n\right)$, and a St\"ormer state as $\left(x_n,x_{n-1}\right)$. In
each of these three cases there exists a unique analytical solution of Eq.
(1) to which these states belong although the corresponding exact
trajectories are not identical.

\subsection{The Order of Approximation of The Leapfrog Scheme}
This very basic and apparently simple question appears to be very confused
and persistently misinterpreted in the literature. Fundamental textbooks
always demonstrate that algorithm (2) is accurate to fourth order but add
that velocities are usually computed with a lower accuracy. Equivalence
between all integrators of this group also appears only when low-order
approximations are used for velocities \cite{Rodger:89}.  At the same time,
Eqs. (3a,b) and the velocity Verlet and Beeman algorithms are only
$O\left(h^3\right)$ accurate. In various numerical tests all these
integrators exhibit an $O\left(h^2\right)$ growth of global errors in
agreement with $O\left( h^3\right)$ order of the local truncation error
\cite{Allen:87,Haile:92}.  This occurs even when no lower-order
approximations are involved, for instance, if Eq. (2) is used and the
deviation of coordinates from an analytical solution is checked
\cite{Venneri:87}. Although these observations are not contradictory they are
rather perplexing, but, however surprising, they are essentially ignored in
the literature. The confusion clearly results from approximations of
velocities which are not included in Eq. (2), but since all these algorithms
can produce the same trajectory the difficulty is only formal and seems to
present no practical interest. We will now see that this reasoning involves a
conceptual error with important consequences.

Consider the standard derivation of Eq. (2), which starts from two Taylor
series from time $t$ in opposite directions
{\mathletters\begin{eqnarray}
x\left(t+h\right)=x\left(t\right)+\dot x\left(t\right)h+\frac 12\ddot x
\left(t\right)h^2+\frac 16\dot{\ddot x}\left(t\right)h^3+O\left(h^4\right)\\
x\left(t-h\right)=x\left(t\right)-\dot x\left(t\right)h+\frac 12\ddot
x\left(t\right)h^2-\frac 16\dot{\ddot x}\left(t\right)h^3+O\left(h^4\right)
\end{eqnarray}}
Adding these two expansions gives Eq. (2), with all odd-order terms
eliminated, and a resultant truncation error of $O\left(h^4\right)$. Now
let us perform a similar derivation for half-step velocities, which gives
\begin{equation}
v\left(t+\frac h2\right)=2v\left(t-\frac h2\right)-v\left(t-\frac{3h}
2\right)+\ddot v\left(t-\frac h2\right)h^2+O\left(h^4\right)
\end{equation}
and by substituting
\begin{equation}
\ddot v\left(t-\frac h2\right)=\frac 1h\left[\ddot x\left(t\right)-\ddot x\left(
t-h\right)\right]+O\left(h^2\right)
\end{equation}
we get a fourth order finite difference equation
\begin{equation}
v_{n+\frac 12}=2v_{n-\frac 12}-v_{n-\frac 32}+\left(a_{n-1}-a_n\right)h^2
\end{equation}
It is readily seen that any trajectory produced by the leapfrog scheme
(3a,b) satisfies this equation, so it is an exact analog of the St\"ormer
formula (2) for velocities. We see, therefore, that the order of
approximation of the leapfrog scheme is four rather than three, both for
coordinates and velocities. Eqs. (2) and (11) are common fourth order
predictors which use values from several time steps to reduce the truncation
error. The fact that the same trajectory can be generated by lower-order
Eqs. (3a,b) results from a fortunate cancelation of errors.

Note, however, that Eqs. (2), (11) and Eqs. (3a,b) are not equivalent. Eq.
(11) provides a leapfrog solution only if initial conditions match Eq. (3a).
Note also, that Eqs. (2) and (11) are coupled only via accelerations - no
relationship between coordinates and velocities is implied {\em a priori.}
Therefore, starting from arbitrary initial conditions a completely
non-leapfrog trajectory may be produced. The relationship between the
general solutions of Eqs. (2), (11) and (3a,b) presents an interesting
question, but it is beyond the scope of the present paper.

\subsection{Accumulation of Errors Along A Leapfrog Trajectory}
The truncation error is in fact only a part of the story because single-step
errors can accumulate thus producing so called global errors
\cite{Allen:87,Haile:92}, which are relevant for assessment of an algorithm
and which, as noted above, exhibit a characteristic $O\left(h^2\right)$
growth. Let us check how, in the case of a leapfrog trajectory, truncation
errors accumulate for conventional testing procedures.

Suppose we have a trajectory of duration $T$; we compute some
parameter $Q\left[x\left(t_i\right),v\left(t_i\right)\right]$ at $M$
time steps $i=1,...,M$ and evaluate
\begin{equation}
\overline {\Delta Q^2}=\frac 1M\sum_{i=1}^M\Delta Q_i^2=\frac 1M\sum_{i=1}^M\left[
Q\left(t_i\right)-Q_0^a\left(t_i\right)\right]^2
\end{equation}
Here $Q_0^a\left(t_i\right)$ is the corresponding analytical value
computed for an exact trajectory starting from the initial state. An overbar
here and below denotes time averaging. Now we have
\begin{equation}
\Delta Q_i=\sum_{j=0}^i\Delta \Delta Q_j
\end{equation}
where
\begin{eqnarray}
\Delta \Delta Q_j=&\Delta Q_{j+1}-\Delta Q_j=Q\left(t_{j+1}\right)-Q\left(
t_j\right)-Q_0^a\left(t_{j+1}\right)+Q_0^a\left(t_j\right)=\nonumber\\
&\left[Q\left(t_{j+1}\right)-Q_j^a\left(t_{j+1}\right)\right]+\left[
Q_j^a\left(t_{j+1}\right)-Q\left(t_j\right)\right]-\left[Q_0^a\left(
t_{j+1}\right)-Q_0^a\left(t_j\right)\right]
\end{eqnarray}
where $Q_j^a\left(t\right)$ refers to the analytical trajectory starting
from jth state in the numerical trajectory. Here only the first bracket in
the r.h.s. is contributed by the local truncation error while the rest
results from deviations of analytical solutions corresponding to different
initial conditions. Discarding higher order terms one can rewrite Eq. (14) as
\begin{equation}
\Delta \Delta Q_j\approx\alpha _jh^m+\left[\dot Q_j^a\left(t_j\right)-\dot
Q_0^a\left(t_j\right)\right]h
\end{equation}
where $\alpha _j$ are constant coefficients and, for the leapfrog scheme, $
m=4$.

In the general case both terms in the r.h.s. of Eq. (15) must be taken into
consideration. Let us check the case when $Q$ is just the coordinate and so
Eq.(12) estimates deviation from an analytical trajectory. Let $x_j^a\left(
t\right)$ and $x_{j+1}^a\left(t\right)$ be two analytical solutions of
Eq. (1) with boundary conditions
{\mathletters\begin{eqnarray}
x_j^a\left(t-h\right)=x_{j-1};\qquad x_j^a\left(t\right)=x_j\\
x_{j+1}^a\left(t\right)=x_j;\qquad x_{j+1}^a\left(t+h\right)=x_{j+1}
\end{eqnarray}}
where $x_{j-1}$, $x_j$ and $x_{j+1}$ are successive points on a numerical
trajectory. We have
{\mathletters\begin{eqnarray}
x_{j+1}^a\left(t+h\right)=x_j+\dot x_{j+1}^a\left(t\right)h+f\left(
x_j\right)\frac{h^2}2+\dot{\ddot x}_{j+1}^a\left(t\right)\frac{h^3}6+\ddot{
\ddot x}_{j+1}^a\left(t\right)\frac{h^4}{24}+O\left(h^5\right)\\
x_j^a\left(t-h\right)=x_j-\dot x_j^a\left(t\right)h+f\left(x_j\right)\frac{
h^2}2-\dot{\ddot x}_j^a\left(t\right)\frac{h^3}6+\ddot{\ddot x}_j^a\left(
t\right)\frac{h^4}{24}+O\left(h^5\right)
\end{eqnarray}}
Taking into account that $x_{j-1}$, $x_j$ and $x_{j+1}$ are related by Eq.
(2) summation of Eqs (17a,b) yields
\begin{equation}
\left[\dot x_{j+1}^a\left(t\right)-\dot x_j^a\left(t\right)\right]h+\left[
\dot{\ddot x}_{j+1}^a\left(t\right)-\dot{\ddot x}_j^a\left(t\right)\right]
\frac{h^3}6+\left[\ddot{\ddot x}_{j+1}^a\left(t\right)+\ddot{\ddot x}_j^a\left(
t\right)\right]\frac{h^4}{24}+O\left(h^5\right)=0
\end{equation}
Thus, except for the special case $\dot x_{j+1}^a\left(t\right)\equiv
\dot x_j^a\left(t\right)$ we have
\begin{equation}
\dot x_{j+1}^a\left(t\right)-\dot x_j^a\left(t\right)=O\left(h^3\right)
\end{equation}
Assuming that $x^a\left(t\right)$ are smooth enough, relation (19) can be
continued over a finite time interval and we may write
\begin{equation}
\dot x_j^a\left(t_j\right)-\dot x_0^a\left(t_j\right)=\sum_{k=0}^j\left[
\dot x_{k+1}^a\left(t_j\right)-\dot x_k^a\left(t_j\right)\right]\approx
\sum_{k=0}^j\beta _kh^3=t_j\beta \left(t_j\right)h^2
\end{equation}
where $\beta _k$ are constant coefficients and $\beta \left(t_j\right)
$ results from averaging over interval $\left(0,t_j\right)$ for a
sufficiently small step size. Thus in Eq. (15) the second term appears to be
$O\left(h^3\right)$ and it dominates over the truncation error given by
the first term. Now, by performing the summation in Eq.(13) similarly to Eq.
(20), we see that the deviation of the leapfrog trajectory from an
analytical solution must be of the order of $O\left(h^2\right)$ in agreement
with usual conclusions \cite{Venneri:87}. We see, therefore, that the square
growth of the deviation of the coordinates from an analytical trajectory has
a more complicated origin than the simple accumulation of truncation errors
and it does not contradict the $O\left(h^4\right)$ truncation error of the
algorithm.

Now let us consider the case when $Q$ is the total energy. For any
analytical solution its time derivative is zero and $\Delta \Delta E_j$ in
Eq. (15) appears to be $O\left(h^4\right)$. The same result is obtained
with the average total energy instead of the analytical value in Eq. (12).
Summation in Eq. (13) now gives the global error on the order of $O\left(
h^3\right)$, that is one order higher than the common conclusion
\cite{Allen:87,vanGunsteren:77}.
There is one complication, however, which was missed in the above
derivations. We tentatively assumed in Eq. (14) that for any state on a
numerical trajectory one can find a reference analytical solution passing
through this state. For the next-step total energy to deviate as $O\left(
h^4\right)$ both coordinates and velocities must be within $O\left(
h^4\right)$ of this reference trajectory, which, in turn, requires that it
pass through both $(x_n,x_{n-1})$ and $(v_{n-1/2},v_{n-3/2})$ according to
Eqs. (2) and (11), respectively. However, since Eq. (1) is only of second
order, such an analytical solution does not necessarily exist. In the
general case, we have two different analytical trajectories corresponding to
coordinates and velocities and the overall single-step error in the total
energy has a contribution from incoherence between these trajectories. On
the other hand, if we use as a reference an analytical trajectory passing
through a leapfrog state or a velocity Verlet state we obtain $O\left(
h^3\right)$ single-step deviation and $O\left(h^2\right)$ global error,
but this is only an upper estimate because it does not take into account the
cancelation of local errors which allows solutions of Eqs. (3a,b) to fit
higher order equations (2) and (11). We conclude, therefore, that the global
error in the total energy for a leapfrog trajectory must be between $O\left(
h^2\right)$ and $O\left(h^3\right)$, but it may depend upon the specific
properties of Eq. (1). That is why the order of global errors in this case
cannot be estimated analytically {\em a priori}, and should rather be measured
numerically. We will see in the next section, however, that this appears
difficult.

\subsection{Effects of Interpolations of Kinetic Energy}
\label{interp}
We are going to show here that the common approach to testing energy
conservation in molecular dynamics in the case of the leapfrog scheme in
fact fails to evaluate the true error of the algorithm because of a
dominating contribution of additional interpolations necessary to compute
on-step kinetic energies. Following many previous studies, we first consider
a simple harmonic oscillator as a model case suitable for analytical
treatment. This simple model appears to have a significant predictive power
because the fastest motions in real systems are nearly harmonic. We will
check this on a more realistic example of protein dynamics.

Consider an oscillator with the Hamiltonian
\begin{equation}
H=\frac 12\left(\dot x^2+\omega ^2x^2\right)
\end{equation}
and the leapfrog equations of motion
{\mathletters\begin{eqnarray}
v_{n+\frac 12}-v_{n-\frac 12}=-\omega ^2x_nh\\
x_{n+1}-x_n=v_{n+\frac 12}h
\end{eqnarray}}
It is known that these finite difference equations have an analog of the
total energy \cite{Hockney:81}. By multiplying Eq. (22a) by
$(v_{n+1/2}+v_{n-1/2})$ with simple algebra one obtains
\begin{equation}
\frac 12\left(v_{n+\frac 12}^2+\omega ^2x_nx_{n+1}\right)=\frac 12\left(
v_{n-\frac 12}^2+\omega ^2x_{n-1}x_n\right)=\varepsilon _1
\end{equation}
Similarly, by multiplying Eq (22b) by $(x_{n+1}+x_n)$ we obtain
\begin{equation}
\frac 12\left(v_{n-\frac 12}v_{n+\frac 12}+\omega ^2x_n^2\right)=\frac
12\left(v_{n-\frac 32}v_{n-\frac 12}+\omega ^2x_{n-1}^2\right)=\varepsilon
_2
\end{equation}
and it is easy to check that $\varepsilon _1=\varepsilon _2=\varepsilon $.
Thus, there is a quantity with the dimension of energy which is exactly
conserved along the numerical trajectory. In the general case a numerical
``first integral'' analogous to $\varepsilon $ does not exist, but it is
very useful in analytical derivations and gives the possibility to reveal
certain qualitative features intrinsic in the common approaches to the
estimation of the on-step total energy. Consider the most common
approximation of the on-step total energy for the leapfrog scheme
\begin{equation}
E_{lf}^n=\frac 12\left[\frac 12\left(v_{n-\frac 12}^2+v_{n+\frac
12}^2\right)+\omega ^2x_n^2\right]=\varepsilon +\frac{\tau ^2}2U_n
\end{equation}
where $\tau =\omega h$ is the reduced step size. Here we made a substitution
of Eqs. (22a) and (24) and denoted potential energy as $U_n$. As expected,
the total energy is not constant and it fluctuates with a small amplitude on
the order of $O\left(\tau ^2\right)$. The relative fluctuation, sometimes
used as a convenient indicator of the accuracy of the trajectory, is simply
\begin{equation}
q_{lf}=\frac{D\left[E_{lf}\right]}{D\left[U\right]}=\frac{\tau ^2}2
\end{equation}
where $D\left[{}\ \right]$ denotes the operator of variance. The fluctuating
term in Eq. (25) is usually used to check for the accuracy and stability of
a computed trajectory. One can note, however, that the analytical form of
this term is somewhat suspicious. Namely, it could have been expected that a
numerical fluctuation of an analytically constant value has a more
complicated form than just a scaled oscillation of the potential energy. In
order to make clear the origin of this oscillation let us consider the
related value for an exact analytical trajectory of an oscillator with
$x\left(0\right)=0$ and $v\left(0\right)=\omega$. We have
\begin{equation}
E_{lf}^a=\frac{\omega ^2}2\left\{\frac 12\left[\cos ^2\omega \left(
t-\frac h2\right)+\cos ^2\omega \left(t+\frac h2\right)\right]+\sin
^2\omega t\right\}=E_a\left(1-\frac{\tau ^2}4\right)+\frac{\tau ^2}
2U\left(t\right)+O\left(h^4\right)
\end{equation}
where $E_a$ and $U\left(t\right)$ are analytical total and potential
energies, respectively. Comparison of this result with Eq. (25) shows that
the $O\left(\tau ^2\right)$ oscillation is exactly same for the numerical
trajectory as for the analytical one suggesting that it is introduced by the
interpolation and has no relation to the accuracy of the trajectory. In
order to validate this suggestion let us check what one gets when more
accurate interpolation formulas are employed. Consider the interpolation
\begin{equation}
K_n=\frac 18\left(3K_{n+\frac 12}+6K_{n-\frac 12}-K_{n-\frac 32}\right)
+O\left(h^3\right)
\end{equation}
where $K_{n-1/2}$ denotes the half-step kinetic energy. It is not difficult
to check that when applied to an exact trajectory it gives an $O\left(\tau
^3\right)$ oscillation around the correct total energy. For the leapfrog
solution, derivations similar to that used for Eq. (25) result in
\begin{equation}
E_3^n=\varepsilon \left(1+\frac{\tau ^2}4\right)+\frac{\tau ^3}2\omega
x_nv_{n-\frac 12}-\frac{\tau ^4}8K_{n-\frac 12}
\end{equation}
Again we see that the amplitude of the oscillation scales as the order of
the interpolation. Note that its phase is shifted by $\pi /2$ from that of $
E_{lf}$. For the following discussion we need one more interpolation
\begin{equation}
K_n=\frac 1{48}\left(15K_{n+\frac 12}+45K_{n-\frac 12}-15K_{n-\frac
32}+3K_{n-\frac 52}\right)+O\left(h^4\right)
\end{equation}
which gives
\begin{equation}
E_4^n=\varepsilon \left(1+\frac{\tau ^2}2\right)+\frac{\tau ^4}{16}\left(
5K_{n-\frac 12}+K_{n-\frac 32}-4U_{n-1}\right)
\end{equation}
with $O\left(\tau ^4\right)$ oscillation, as expected. $E_4$ oscillates in
phase with the kinetic rather than with potential energy, which means that
it is shifted by $\pi $ from $E_{lf}$.

One might conclude that in this special case the total energy is conserved
exactly with any time step size. We will later see, however, that actually
the leapfrog trajectory deviates from an exact constant energy hypersurface,
but these deviations are regular and they appear to be compensated by the
approximation errors of the interpolation formulas. This occasional
cancelation certainly results from the fundamental solutions of the finite
difference equation (2) being sine and cosine functions \cite{Hockney:81},
but it is
rather instructive because it demonstrates that interpolations can produce
very misleading effects. The most interesting for us, however, is the form
of the oscillations produced by the interpolation formulas. We will now see
that qualitatively similar behavior is observed in real simulations.

Figs. 1-3 present a 50 fsec interval of a molecular dynamics trajectory
computed with three different time steps: 0.5 fsec, 0.1 fsec and 0.05 fsec,
respectively. The system modeled consists of an immunoglobulin binding
domain of streptococcal protein G \cite{pgb:} which is a 56 residue $\alpha $/$
\beta $ protein subunit (file 1pgb in the Protein Database \cite{PDB:}) with 24
bound water molecules presented in the crystal structure. All hydrogens are
considered explicitly thus giving 927 atoms. The trajectory was computed
without constraints upon the protein structure and with TIP3P water
molecules held rigid. Other details of the simulation protocol are given
in Section \ref{tests} below. In such a system the fastest oscillations are
due to bond
stretching of hydrogens with a period of about 10 fsec, which is clear in
the time profiles of the potential energy shown in Figs. 1a-3a. Figures
1b-3b show the time profile of the total energy computed by Eq. (25). Note
that according to Eqs. (25), (29) and (31) the amplitude of the oscillation
of the total energy grows with the frequency, which means that, in systems
of many oscillators, interpolations tend to filter lower frequencies. It is
not surprising therefore that in Figs. 1b-3b the highest frequency strongly
dominates compared with Figs. 1a-3a. Note, however, that in all three
figures $E_{lf}$ oscillation has exactly the same phase as that of the
potential energy.

The profiles of the potential energy in Figs. 1a - 3a are indistinguishable,
which means that with a step size of 0.5 fsec the trajectory is perfectly
accurate. In agreement with the simple case considered above the shapes of
the profiles of the total energy in Figs. 1b-3b also appear to be
indistinguishable, with the amplitude of the oscillation scaled as the
square of the step size. The identity of the three profiles in this case is
not evident {\em a priori} and it is difficult to explain unless the above
observations for an oscillator are not invoked. Thus, despite the simplicity
of the model employed, Eq. (25) very accurately predicts both the growth
rate with the step size and the shape of the fluctuation of the total energy
in a much more complex system.

Now consider the time profiles of the total energy computed with higher
order interpolations. Figs. 1c and 1d show that both $E_3$ and $E_4$ behave
similarly to $E_{lf}$ and as predicted by Eqs (29) and (31). Note that the
oscillations of $E_3$ and $E_4$ are shifted by $\pi $/2 and $\pi $,
respectively, from $E_{lf}$ in Fig. 1b. With a reduced step size, however,
both $E_3$ and $E_4$ reveal some new features. Similarly to $E_{lf}$ the $
E_3 $ oscillation in Figs. 2c and 3c remains dominated by high-frequency
oscillation, although its overall profile slightly changes. It is easy to
see that the amplitude of the high-frequency oscillation scales as $O\left(
h^3\right)$ and that the change in the profile is likely to be due to
an underlying fluctuation which scales slower than $O\left(h^3\right)$
. Finally, for $E_4$ oscillation in Figs. 2d and 3d we observe a profile of
the fluctuation qualitatively different from those in Figs. 1(a-d). The
amplitude of the oscillation in Fig. 1d reduced by factors of 625 and 10$^4$
in Figs. 2d and 3d, respectively, would present a negligible part of the
remaining fluctuation and that is why this hidden profile is revealed. This
residual deviation of the total energy can be attributed to the
interpolation-free error of the trajectory itself. Comparison of Figs. 2d
and 3d shows that the amplitude of the remaining fluctuation scales faster
than $O\left(h^2\right)$ but slower than $O\left(h^3\right)$, as
we expected. If scaled back to the step size of 0.5 fsec used in Fig.1 this
fluctuation can account for only a few percent of the whole amplitude, which
indicates that the total energy is really much better conserved than it
appears to be. It is interesting that the high-frequency harmonic
oscillation that dominates in Figs.1 (b-d) essentially disappears in Figs.
2d and 3d suggesting that harmonic motions in the system really do not
contribute to the apparent fluctuation of the total energy, which is
produced by slower anharmonic motions.

Let us now summarize the above results. In the case of an oscillator,
fluctuations are produced by interpolations only, their phases and
amplitudes are related to those of the kinetic and potential energies in a
simple way. For a realistic microcanonical ensemble with all common types of
interactions, the total energy behaves similarly for the second and third
order interpolations, as well as with the fourth order one at a moderately
small step size. For an extremely small step size, however, a qualitative
difference is observed between the fourth and lower order interpolations.
All this agrees very well with the global error in the total energy between $
O\left(\tau ^2\right)$ and $O\left(\tau ^3\right)$ corresponding to $
O\left(\tau ^4\right)$ local error in the trajectory. Although the
trajectory contribution can be revealed by interpolations of the fourth and
higher orders, with practical time step values, the interpolation errors
always dominate overwhelmingly. Taking into account that, because of the
rapid growth of higher derivatives on repulsive walls of non- bonded
potentials, higher order interpolations tend to be less accurate in ``less
harmonic'' systems, we are obliged to conclude that straightforward
interpolations in principle cannot adequately evaluate instantaneous on-step
velocities and kinetic energies in the case of the leapfrog scheme. This
approach, therefore, should not be used for assessing the accuracy of the
algorithm.

\subsection{Alternative Strategy for Checking Energy Conservation}
\label{tests}
In this section we consider how numerical tests with microcanonical
ensembles can be arranged so that the real accuracy of conservation of the
total energy can be assessed. The main idea is simple. Note that all
interpolations of the kinetic energy such as Eqs. (28) and (30) give the same
average kinetic energy. For any sufficiently long analytical trajectory these
interpolations, upon averaging, result in a correct average kinetic energy
and, consequently, exact total energy. This property does not depend upon the
specific form of the Hamiltonian or the number of degrees of freedom. Note,
for example, that in Figs. 1-3 with the same step size, $E_{lf}$, $E_3$ and
$E_4$ oscillate around the same average. At the same time there is a
distinguishable difference between the averages for the three given values of
the step size. We see, therefore, that unlike instantaneous energies, their
time averages appear to be free from interpolation errors and they represent
adequate indicators of the accuracy of the energy conservation.

Suppose we know the analytical value of the total energy. We could then
repeatedly calculate a long enough test trajectory with gradually growing
time steps and compare the average total energy with the analytical one. As
long as these two values are close to each other, the computed trajectory
remains close to a correct hypersurface in phase space. Instead of the
unknown analytical total energy, we can use that obtained for the same
trajectory with a very small step size. It is important to make sure,
however, that the test trajectory starts from the same state on a fixed
constant-energy hypersurface with each step size, which is not automatic in
the case of the leapfrog scheme. Let us consider the implementation of this
testing procedure for a harmonic oscillator.

By using Eqs. (22a) and (24) one can derive an identity
\begin{equation}
U_n\equiv \frac \varepsilon 2+\frac 12\left[U_n-\frac 12\left(K_{n-\frac
12}+K_{n+\frac 12}\right)\right]+\frac{\tau ^4}4U_n
\end{equation}
Denoting
\begin{equation}
\delta _n=U_n-\frac 12\left(K_{n-\frac 12}+K_{n+\frac 12}\right)
\end{equation}
we obtain
\begin{equation}
U_n=\frac 1{2\left(1-\frac{\tau ^2}4\right)}\left(\varepsilon +\delta
_n\right)
\end{equation}
It can be shown by a straightforward solution of the finite difference
equations (22a,b), that within the time step range of stability of an
oscillator $\bar U=\bar K$. Therefore $\delta _n$ presents an oscillation
around a zero average and we have
\begin{equation}
\bar E_{lf}=2\bar U=\frac \varepsilon {\left(1-\frac{\tau ^2}4\right)}
\end{equation}
Let us now check how accurately the numerical average total energy
approximates exact values, that is total energies corresponding to
constant-energy hypersurfaces sampled by a leapfrog trajectory. Consider an
analytical trajectory passing through a leapfrog state $\left(
v_{n-1/2},x_n\right).$ We have
\begin{equation}
\varepsilon =E_a\left\{\sin ^2\omega t+\cos \omega \left(t-\frac h2\right)
\left[\cos \omega \left(t-\frac h2\right)-\tau \sin \omega t\right]
\right\}
\end{equation}
From Eqs. (35) and (36) by using a Taylor series expansion, we obtain
\begin{equation}
\frac{E_a}{\bar E_{lf}}=1+\left[\frac{\tau ^3}{48}+O(\tau ^5)\right]\sin
2\omega t-\left[\frac{\tau ^4}{96}+O(\tau ^6)\right]\cos 2\omega t-\frac{
\tau ^4}{96}-\frac{\tau ^6}{64}+O(\tau ^8)
\end{equation}
The r.h.s. of Eq. (37) involves two qualitatively different types of
deviations. The first depends upon the current phase of the oscillation and
it in fact presents the true measure of the energy conservation along the
trajectory: as long as the phase-dependent terms in Eq. (37) are small, all
states on the trajectory belong, or are close, to the same exact constant
energy hypersurface. When these terms become too large, successive leapfrog
states correspond to different hypersurfaces, but the corresponding
analytical total energies oscillate around a certain value. The second type
of deviation is phase-independent and it is presented by the fourth and
sixth-order terms. Because of this deviation, with large step size all
leapfrog states appear to belong to hypersurfaces of a systematically lower
energy than the computed average value, which means that they are no longer
sampled ergodically. The magnitudes of both these types of deviations can be
easily evaluated, and it turns out that for $\tau\lesssim 1.1$ the
fluctuating part dominates, while above this level the phase-independent
contribution rapidly becomes overwhelming due to its sixth order growth.

Now consider what one would observe in the numerical tests outlined above.
It can be seen from Eq. (35) that by simply increasing the step size one
obtains an $O(\tau )$ regular growth of $\bar U$ and $\bar E_{lf}$. It is clear,
however, that, in this case, the starting leapfrog state in phase space is
effectively moved from one constant energy hypersurface to another, which
results in a regular drift of averages. In order to remove this drift, the
trajectory should start from the same hypersurface, that is from $x_0$ and
$v_{-1/2}$ corresponding to the same analytical trajectory. It is easy to see
that this case is also described by Eq. (37) if it is read in an opposite
sense. Namely, now $E_a$ and the phase of the oscillating terms on the right
are constant because they correspond to the initial constant energy
hypersurface and the phase of the starting state. Equation (37) therefore
describes the step size dependence of $E_{lf}$ which beyond $\tau\approx
1.1$\ should rapidly deviate upward from its zero step size limit.
It turns out that this estimate holds quite well for representative molecular
models as well, which is illustrated by Fig. 4.

This figure presents an example of application of the proposed test to
protein dynamics. The simulations were made by AMBER molecular modeling
program \cite{AMBER:} with AMBER94 force field \cite{AMBER94:}. The system
considered was same as in Figs.1-3, that is with only water bond lengths and
bond angles constrained by the SHAKE algorithm\cite{SHAKE:}. Initial data
for these tests were prepared as follows. The equilibration was more or less
standard and included minimization of the crystal structure followed by a
12.5 psec run starting from Maxwell distribution at 300K with periodic
temperature control and a step size of 0.5 fsec. After that the step size was
reduced to 0.25 fsec and a short trajectory of 150 fsec was calculated, with
the final part stored and used in place of an analytical trajectory for
generating initial data. The last point was used as the starting state for a
10 psec test trajectory which was computed with gradually growing step sizes
and half-step velocities taken at appropriate time intervals from
coordinates.

In order to apply our test to a real molecular system we should also take
into account that, since in common empirical potentials the absolute value of
the potential energy involves many non-harmonic contributions which always
remain far from zero, it is no longer sensible to compare the deviations with
the absolute energy values. An appropriate natural scale, however, is given
by the variance of the instantaneous potential energy over the test
trajectory. One may reasonably consider for $\bar U$ the deviation of
$\pm 0.1D\left[U\right]$, for instance, as an acceptable level of
accuracy. Note that in the oscillator case this gives approximately same
criterion as above. The deviation of $E_{lf}$ should be two times larger because
it involves a similar contribution from the kinetic energy.

Figure 4 (a) shows the step size dependence of $D\left[U\right]$.  The dotted
line in this figure shows the low step size level used to plot the
corresponding bands of acceptable accuracy in Figs. 4 (b) and (c).  As seen
in Fig. 4(b) the fluctuation of $\bar U$ remains within the band $\pm
0.1D\left[U\right]$ up to a time step of 1.7 fsec.  These fluctuations are
relatively large because they are affected by rare conformational transitions
which cannot be averaged during a test trajectory. Oppositely, $E_{lf}$ shown
in Fig. 4 (c) grows steadily, but its deviation also remains within the band
of acceptable accuracy up to the same step size. Assuming that this value
corresponds to $\tau\approx 1.1$ one obtains a frequency of 3400 cm$^{-1}$,
that is exactly that of bond stretching of peptide hydrogens. The whole
spectrum of bond stretching modes of hydrogens covers the range 3000--3700
cm$^{-1}$, which corresponds to characteristic step sizes in the range
1.6--1.9 fsec. We may conclude, therefore, that Fig. 4 demonstrates a
good agreement with the single oscillator model considered above.

At this point it is convenient to compare our step size estimates with the
common recommendations.  Although no strict rules exist, it is usually
considered that the computed trajectory provides an acceptably accurate
sampling when the relative fluctuation $q_{lf}$ given by Eq. (26) is less
than 0.1, which gives $\tau\lesssim 0.4$ , i.e. at least 14 time steps per a
single cycle of the oscillation. Our estimates show that actually this level
of accuracy is reached with an almost three times larger time step, although
in this case $q_{lf}$ is already so large that the energy conservation seems
to be lost.  In practice, however, the above recommendations are almost never
respected and most often molecular dynamics trajectories are computed with
less than 10, or even less than 5 steps, per the shortest cycle
\cite{Levitt:83}. This practice thus appears to be well justified and it is
clear from Fig. 4 that the leapfrog trajectory really manages to hold to a
constant energy hypersurface with somewhat larger time steps than commonly
recommended.

\subsection{Leapfrog Scheme Versus Its Relatives}
In this section we will demonstrate that, contrary to the conventional point
of view the leapfrog scheme exhibits exceptional properties compared with the
other algorithms of this group.

The results obtained in Section \ref{interp} above obviously are not
applicable to the velocity Verlet and Beeman algorithms, Eqs. (4-5). In these
cases, interpolations appear to be built into the algorithms, which makes
them only $O\left(h^3\right)$ accurate in velocities. (Note that
interpolation (7) has the same order of truncation as that in Eq. (6)).
Oscillations of the instantaneous total energy result from the algorithms
themselves and they grow simply as $O\left(h^2\right)$ in agreement with the
order of the truncation error. Let us look more thoroughly at the total
energy computed by the velocity Verlet integrator. Similarly to Eq.(25) we
obtain for the oscillator
\begin{equation}
E_{vv}^n=\frac 12\left[\frac 14\left(v_{n-\frac 12}+v_{n+\frac 12}\right)
^2+\omega ^2x_n^2\right]=\varepsilon +\frac{\tau ^2}4U_n
\end{equation}
We see that $E_{vv}$ behaves similarly to $E_{lf}$ with two times smaller
amplitude of the oscillation and always $E_{vv}<E_{lf}$, which means that, in
the case of velocity Verlet, the temperature estimated for the same
trajectory is somewhat lower. In real simulations $q_{vv}$ is normally
smaller by a factor of 0.5-0.7 and the corresponding temperature can be lower
by as much as a few degrees. The origin of these differences is clear from
the previous discussion and it is not very interesting.

Since $\bar E_{lf}$ and $\bar E_{vv}$ are $O\left(\tau^2\right)$ different it
is clear that $\bar E_{vv}$ gives one order less accurate estimate of the
total energy of the analytical trajectory approximated by leapfrog states.
The same order of approximation is obtained if we consider the trajectories
passing through velocity Verlet states. In this case the
analytical energies are given simply by Eq. (38), and we have
\begin{equation}
E_a-\bar E_{vv}=\frac{\tau ^2}4\left(U_n-\bar U\right) \label{vv}
\end{equation}
which shows that successive states belong to hypersurfaces of $O(\tau ^2)$
different energies. If we consider $E_a$ and $U_n$ in Eq. (\ref{vv}) as
constants and $\bar E_{vv}$ and $\tau $ as variables we obtain the rate of
growth of the deviation of $\bar E_{vv}$ with the step size. We see that it
is on the order of $O(\tau ^2)$, i.e. similar to the leapfrog scheme with
fixed $\varepsilon$ according to Eq. (35). All these derivations can be
readily reproduced for Beeman and the original St\"ormer algorithms which both
deviate from their starting hypersurfaces as $O(\tau^2)$.

The above results can be summarized as follows. Consider a numerical
trajectory of an oscillator in phase space. It may be represented by
St\"ormer-Verlet states $(x_n,x_{n-1})$, velocity Verlet states $(x_n,v_n)$
or by leapfrog states $(x_n,v_{n-1/2})$. According to Eq. (37), the leapfrog
states sample from constant energy hypersurfaces which are $O(\tau^3)$ close
to each other, while in the former two cases a broader $O(\tau^2)$ spectrum
of energies is covered. In this sense one can say that the leapfrog
trajectory presents the most accurate sampling among the three
representations.

\section{Conclusions}
The results presented in the paper shed some light upon the long-standing
confusion involved in the conventional interpretation of the common
algorithms of the classical molecular dynamics. The main practical result is
a new simple testing scheme for energy conservation, which gives a better
estimate of the quality of leapfrog trajectories. It is shown that the
leapfrog trajectories provide accurate sampling in phase space with large
time step values, when the apparent total energy obtained by routine
interpolations is no longer conserved. This explains the recognized quality
of thermodynamic averages at large step size. The new testing scheme should
be particularly useful for studying the time step limitations in various
molecular models appearing in internal coordinate molecular dynamics
simulations of polymers \cite{Mz:unp}.

\begin{figure}\caption{Time dependence of the potential energy (a) and the
total energy (c-d) for an unconstrained protein molecule. All energies are in
kcal/mole.  The molecular dynamics trajectory was computed with the leapfrog
algorithm and a time step of 0.5 fsec. Three different estimates of the total
energy for figures (c)-(d) were obtained with interpolations of second, third
and fourth order, respectively, applied to half-step kinetic energies. The
average total energy subtracted from the instantaneous values in figures
(c)-(d) was 254.68 kcal/mole.}
\end{figure}
\begin{figure}\caption{Same as Fig. 1, but with the molecular dynamics
trajectory computed with a time step of 0.1 fsec. The average total energy in
Figs. (c)-(d) was 255.42 kcal/mole.}
\end{figure}
\begin{figure}\caption{Same as Fig. 1, but with molecular dynamics
trajectory computed with a time step of 0.05 fsec. The average total energy
in Figs. (c)-(d) was 255.51 kcal/mole.}
\end{figure}
\begin{figure}\caption{Time step dependencies of the potential
energy (b), its time variance (a) and the total energy (c) computed over a 10
psec molecular dynamics trajectory. All energies are in kcal/mole.  The
dotted lines in figures (b) and (c) show the corresponding bands of
acceptable deviation defined as described in the text from the low time step
limit of variance indicated by the dotted line in figure (a).}
\end{figure}
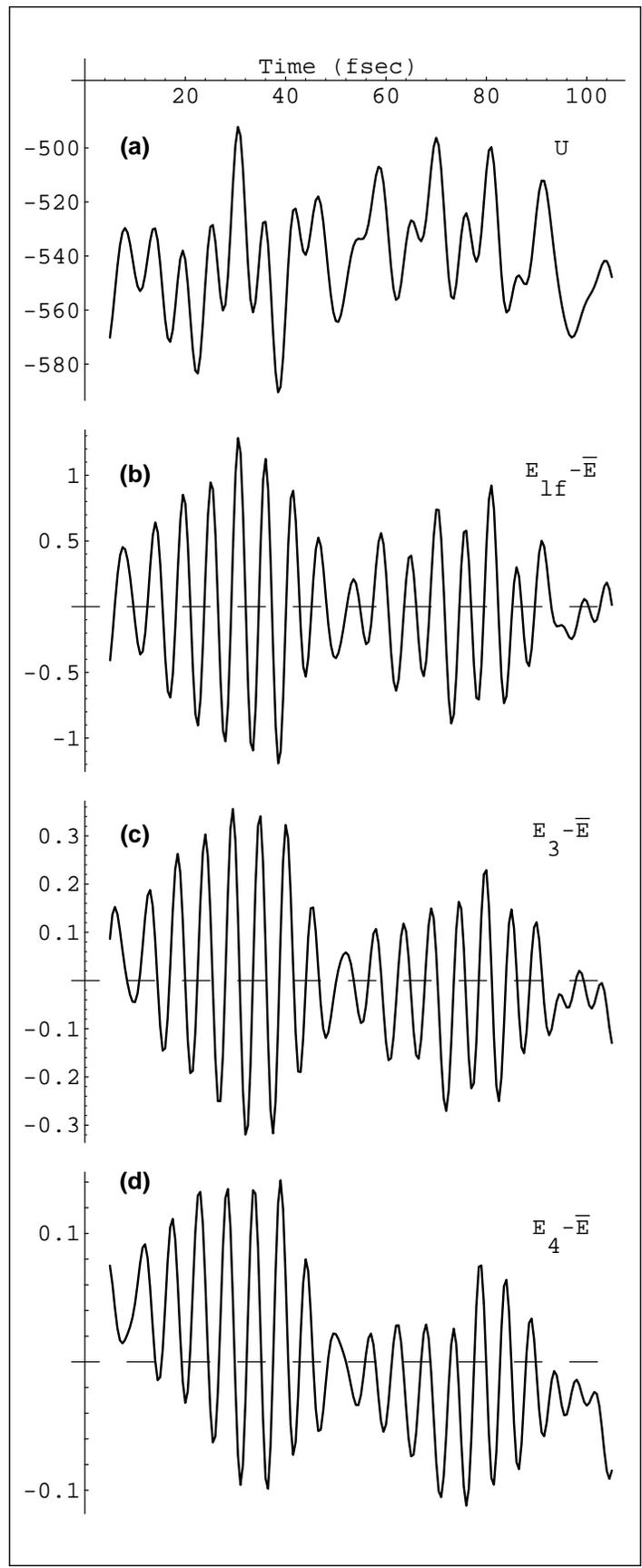
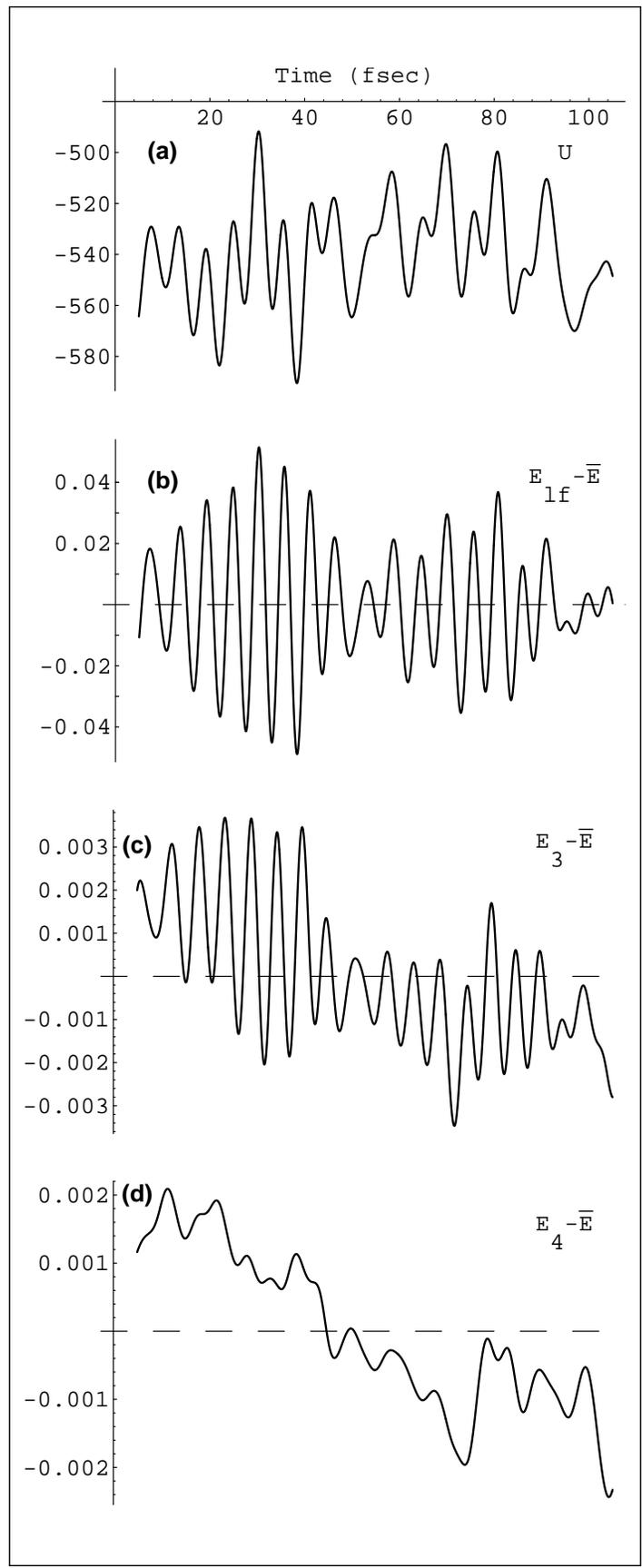
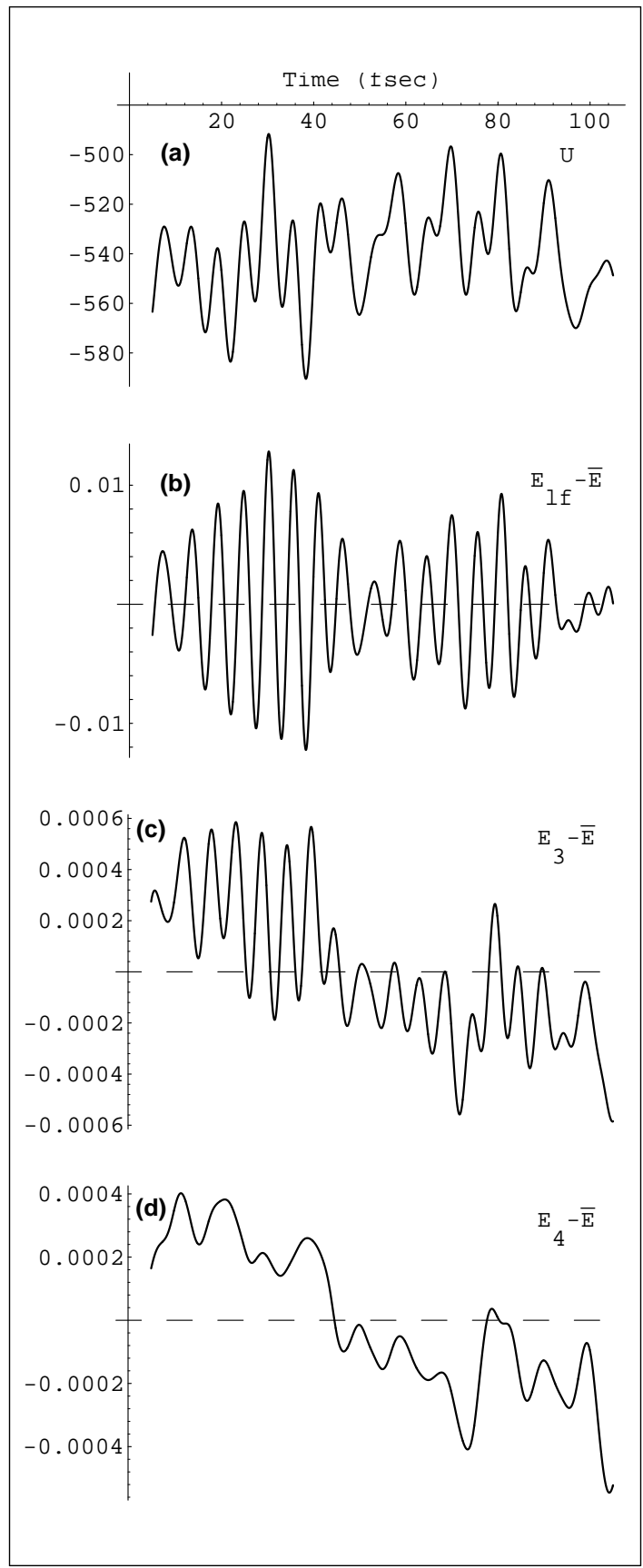
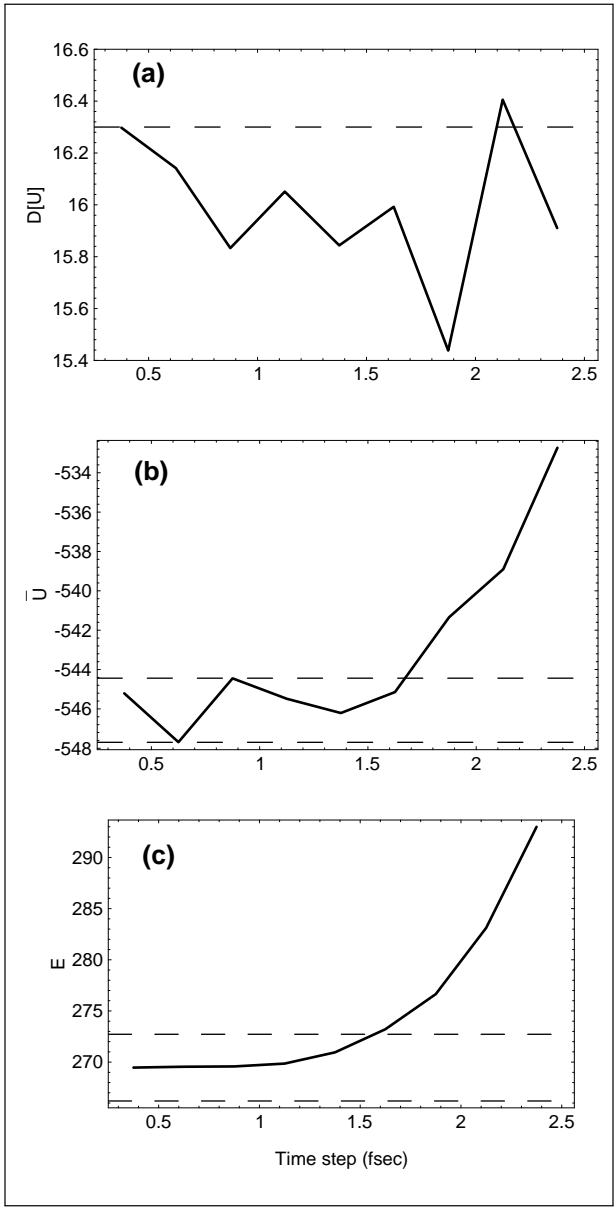
\end{document}